\documentclass{aastex631}

\newcommand\footnoteref[1]

\usepackage{comment}
\usepackage[multiple]{footmisc}
\usepackage{subfigure}
\usepackage[para,flushleft]{threeparttable}
\usepackage{amsmath}

\received{September 16, 2024}
\accepted{January 17, 2025}
\submitjournal{ApJ}

\begin{document}
\title{ngVLA Synthetic Observations of Ionized Gas in Massive Protostars}
\correspondingauthor{Jes\'us Miguel J\'aquez Dom\'inguez}
\email{j.jaquez@irya.unam.mx}

\author[0000-0002-1912-5394]{Jes\'us M. J\'aquez-Dom\'inguez}
\affiliation{Instituto de Radioastronom\'ia y Astrof\'isica, Universidad Nacional Aut\'onoma de M\'exico, Antigua Carretera a P\'atzcuaro \# 8701, Ex-Hda. San Jos\'e de la Huerta, Morelia, Michoac\'an, M\'exico C.P. 58089}

\author[0000-0003-1480-4643]{Roberto Galv\'an-Madrid}

\author[0000-0001-8365-6563]{Alfonso Trejo-Cruz}
\affiliation{Instituto de Radioastronom\'ia y Astrof\'isica, Universidad Nacional Aut\'onoma de M\'exico, Antigua Carretera a P\'atzcuaro \# 8701, Ex-Hda. San Jos\'e de la Huerta, Morelia, Michoac\'an, M\'exico C.P. 58089}

\author[0000-0003-2862-5363]{Carlos Carrasco-González}
\affiliation{Instituto de Radioastronom\'ia y Astrof\'isica, Universidad Nacional Aut\'onoma de M\'exico, Antigua Carretera a P\'atzcuaro \# 8701, Ex-Hda. San Jos\'e de la Huerta, Morelia, Michoac\'an, M\'exico C.P. 58089}

\author[0000-0002-7042-1965]{Jacopo Fritz}
\affiliation{Instituto de Radioastronom\'ia y Astrof\'isica, Universidad Nacional Aut\'onoma de M\'exico, Antigua Carretera a P\'atzcuaro \# 8701, Ex-Hda. San Jos\'e de la Huerta, Morelia, Michoac\'an, M\'exico C.P. 58089}

\author[0000-0002-2260-7677]{Susana Lizano}
\affiliation{Instituto de Radioastronom\'ia y Astrof\'isica, Universidad Nacional Aut\'onoma de M\'exico, Antigua Carretera a P\'atzcuaro \# 8701, Ex-Hda. San Jos\'e de la Huerta, Morelia, Michoac\'an, M\'exico C.P. 58089}

\author[0000-0002-9569-9234]{Aina Palau}
\affiliation{Instituto de Radioastronom\'ia y Astrof\'isica, Universidad Nacional Aut\'onoma de M\'exico, Antigua Carretera a P\'atzcuaro \# 8701, Ex-Hda. San Jos\'e de la Huerta, Morelia, Michoac\'an, M\'exico C.P. 58089}

\author[0000-0001-8446-3026]{Andrés F. Izquierdo}
\affiliation{Department of Astronomy, University of Florida, Gainesville, FL 32611, USA}

\author[0000-0003-2737-5681]{Luis F. Rodr\'iguez}
\affiliation{Instituto de Radioastronom\'ia y Astrof\'isica, Universidad Nacional Aut\'onoma de M\'exico, Antigua Carretera a P\'atzcuaro \# 8701, Ex-Hda. San Jos\'e de la Huerta, Morelia, Michoac\'an, M\'exico C.P. 58089}

\author[0000-0003-1933-4636]{Alice Pasetto}
\affiliation{Instituto de Radioastronom\'ia y Astrof\'isica, Universidad Nacional Aut\'onoma de M\'exico, Antigua Carretera a P\'atzcuaro \# 8701, Ex-Hda. San Jos\'e de la Huerta, Morelia, Michoac\'an, M\'exico C.P. 58089}

\author[0000-0003-4444-5602]{Stanley Kurtz}
\affiliation{Instituto de Radioastronom\'ia y Astrof\'isica, Universidad Nacional Aut\'onoma de M\'exico, Antigua Carretera a P\'atzcuaro \# 8701, Ex-Hda. San Jos\'e de la Huerta, Morelia, Michoac\'an, M\'exico C.P. 58089}

\author{Thomas Peters}
\affiliation{Max-Planck-Institut f\"{u}r Astrophysik, Karl-Schwarzschild-Str. 1, D-85748 Garching, Germany}

\author[0000-0002-2640-5917]{Eric F. Jiménez-Andrade}
\affiliation{Instituto de Radioastronom\'ia y Astrof\'isica, Universidad Nacional Aut\'onoma de M\'exico, Antigua Carretera a P\'atzcuaro \# 8701, Ex-Hda. San Jos\'e de la Huerta, Morelia, Michoac\'an, M\'exico C.P. 58089}

\author[0000-0003-2343-7937]{Luis A. Zapata}
\affiliation{Instituto de Radioastronom\'ia y Astrof\'isica, Universidad Nacional Aut\'onoma de M\'exico, Antigua Carretera a P\'atzcuaro \# 8701, Ex-Hda. San Jos\'e de la Huerta, Morelia, Michoac\'an, M\'exico C.P. 58089}

\begin{abstract}
\noindent
Massive star formation involves significant ionization in the innermost regions near the central object, such as gravitationally trapped \textsc{Hii} regions, jets, ionized disks, or winds. 
Resolved observations of the associated continuum and recombination line emission are crucial for guiding theory. The next-generation Very Large Array (ngVLA) will enable unprecedented observations of thermal emission with 1 mas resolution, providing a new perspective on massive star formation at scales down to a few astronomical units at kiloparsec distances.
This work presents synthetic interferometric ngVLA observations of the free-free continuum (93-GHz band), $\mathrm{H41\alpha}$, and $\mathrm{H38\alpha}$ recombination lines from ionized jets and disks around massive protostars. Using the \texttt{sf3dmodels} Python package, we generate gas distributions based on analytical models, which are then processed through the \texttt{RADMC-3D} radiative transfer code. 
Our results indicate that the ngVLA can easily resolve, both spatially and spectrally, the ionized jet from a 15 $\mathrm{M_\odot}$ protostar at 700 pc, distinguishing between collimated jets and wide-angle winds, and resolving their launching radii, widths, and any substructure down to a few astronomical units. Detailed studies of radio jets launched by massive protostars will be feasible up to distances of $\sim 2$ kpc.
Furthermore, ngVLA will be able to study in detail the ionized disks around massive ($> 10~\mathrm{M_\odot}$) protostars up to distances from 4 to 12 kpc, 
resolving their kinematics and enabling the measurement of their central masses across the Galaxy. 
These observations can be conducted with on-source integrations of only a few hours.
\end{abstract}

\keywords{Protostars (1302) --- Star formation (1569) --- HII regions (694) --- Early stellar evolution (434) --- Radio jets (1347) --- Radio interferometry (1346) --- Stellar accretion disks(1579)}

\section{Introduction} \label{sec:intro}
Massive stars are crucial to galaxy evolution, dominating radiation and shaping the interstellar medium. Their feedback influences star formation; they enrich the cosmos with heavy elements throughout their lifetimes. However, studying these stars is challenging because of their rarity, rapid evolution, and large distances from Earth. High-resolution observations are essential, especially when observing them during their protostellar phases.

Both models and observations indicate that the gas closer to a protostar becomes ionized during the final stages of accretion in massive star formation. 
At a given stage, this ionized gas forms a small ($d \leq 0.03$ pc) and very dense ($n \geq 10^6 \ \mathrm{cm^{-3}}$) \textsc{H\,ii} region, called hypercompact (HC) \citep{Kurtz05}. 
The origin of these regions is still a matter of debate, but many studies suggest that the ionized gas is in the form of gravitationally-trapped disks and envelopes, or due to winds and jets \citep[e.g.,][]{Hollenbach94,Keto2003ApJ...599.1196K,Keto2007ApJ...666..976K,Tanaka2016}. 
Understanding the physical processes related to this ionized gas is vital in the field of star formation, as it will help us to understand the transition stages of a massive protostar toward the main sequence star. 

\begin{figure}[th]
    \centering
    \includegraphics[width = \textwidth]{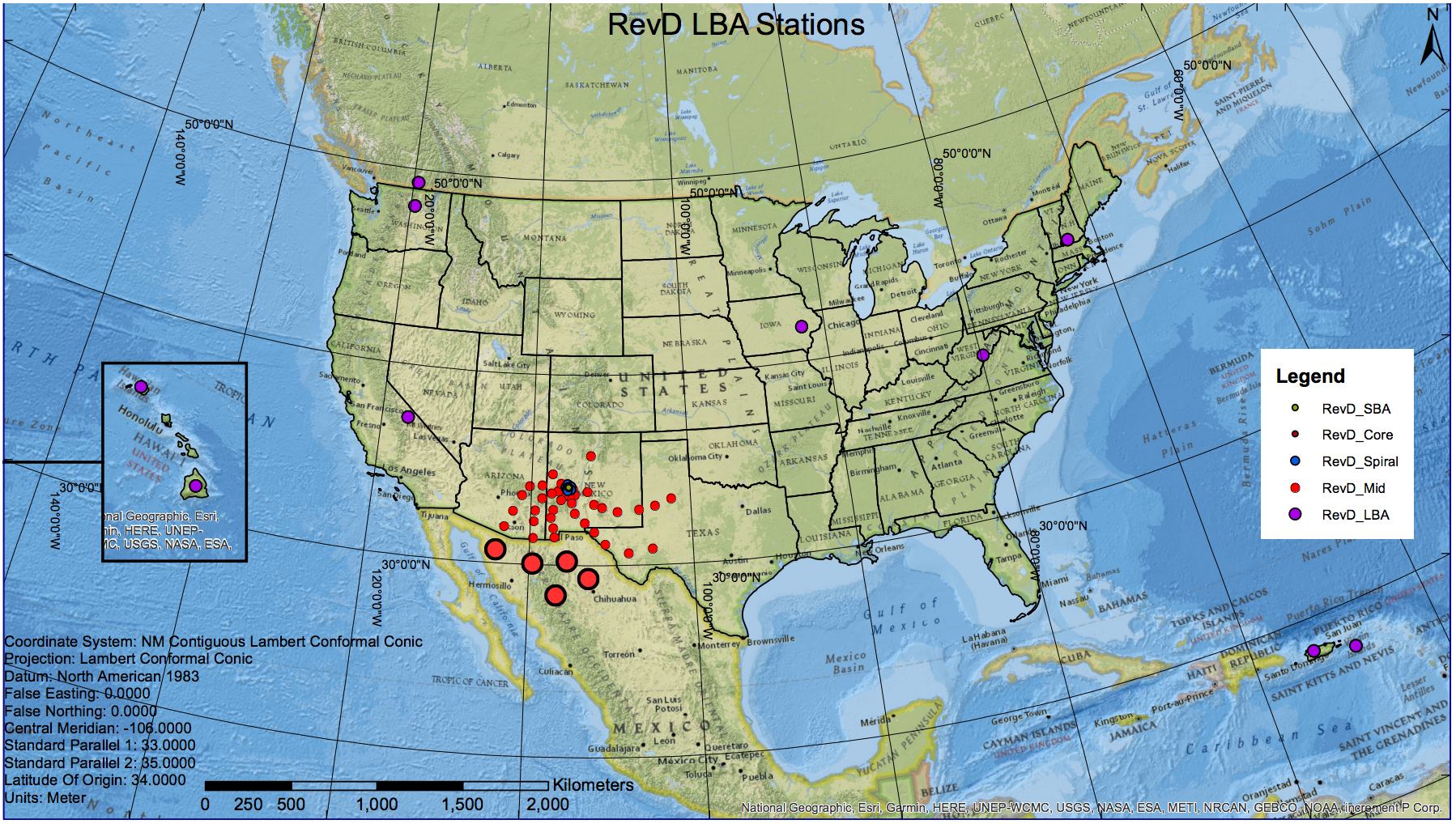}
    \caption{Latest public ngVLA configuration design (Rev D), described in ngVLA memo \#92\footref{revD}. 
The figure shows the mid- (red points) and long-baseline arrays (magenta). The points representing the antennas that will be placed in northern Mexico are magnified for easier recognition.}
    \label{fig:ngVLA_revD_configuration}
\end{figure}

Despite the capabilities of current observatories such as the Atacama Large Millimeter/submillimeter Array (ALMA) and the Karl G. Jansky Very Large Array (VLA), only the closest HC \textsc{H\,ii} regions can be partially resolved in continuum observations, and the brightest among them are sometimes detected in radio recombination lines \citep[RRL, e.g.,][]{van_der_Tak2005A&A...437..947V, DePree2004,GalvanMadrid2009, Sanchez-Monge2013ApJ...766..114S, Guzman14,Rosero2019ApJ...880...99R}.   
In this panorama, the construction of new observatories such as the next generation Very Large Array (ngVLA)\footnote{\url{https://ngvla.nrao.edu}} is a critical step. This future facility is expected to become an indispensable tool for the study of such objects through the Galaxy. 
The ngVLA is being built by the National Radio Astronomy Observatory (NRAO) and is expected to be operational in the early 2030s. The array design is described in \citet{Selina18} and the main scientific objectives are summarized in \citet{Murphy18}. 
The ngVLA will consist of 244 18-meter diameter antennas and 19 6-meter antennas distributed throughout the United States, Mexico, and Canada (e.g., ngVLA memo \#92\footnote{\label{revD}\url{https://library.nrao.edu/public/memos/ngvla/NGVLA\_92.pdf}}). 
In Figure \ref{fig:ngVLA_revD_configuration}, we show the geographical distribution of the antennas. The ngVLA will be made up of 5 configuration components that we list and describe in Table \ref{tab:Rev_D_configuration}. 
This interferometer is designed to operate in the frequency range of 1.2 GHz (21 cm) to 116 GHz (2.6 mm), with up to 20 GHz of instantaneously processed bandwidth.  
We refer the reader to the official documents and memos\footnote{\url{https://ngvla.nrao.edu/page/projdoc}}\textsuperscript{,}\footnote{\url{https://ngvla.nrao.edu/page/memos}} for more technical information on this future facility. 

The sensitivity ($< \mu \mathrm{Jy}$ in continuum observations) and resolution ($\sim$ mas) that this observatory will achieve will allow the spatially and spectrally to resolve the smallest \textsc{H\,ii} regions in massive star-forming regions. 
This will open a new window in the study of the early feedback stages of star formation such as jets, winds, and the transition from gravitationally trapped \textsc{H\,ii} regions to freely expanding regions. With this, our understanding of the physical processes that stop accretion and set a limit to the final stellar masses can be further improved.

\begin{deluxetable*}{c c c c c c }
        \tablecaption{Current reference design (Rev D) ngVLA configuration components. Adapted from ngVLA memo \# 92\footref{revD} \label{tab:Rev_D_configuration}.}
    \tablehead{
         \colhead{Component} &
         \colhead{$N_\mathrm{ants}$\tablenotemark{\scriptsize a}} & \colhead{$D_\mathrm{ants}$\tablenotemark{\scriptsize b}} & \colhead{$B_\mathrm{min}$\tablenotemark{\scriptsize c}} & \colhead{$B_\mathrm{max}$\tablenotemark{\scriptsize d}} & \colhead{Description} \\
          & & \colhead{(meters)} & \colhead{(meters)} & \colhead{(meters)}
          }
         \startdata
         Core   & 114 & 18 & 39  & $4.2\times10^3$ & radial-random at VLA\\
         Spiral & 54  & 18 & 810 & $3.9\times10^4$ & 5 arm exponential spiral at VLA \\
         Mid    & 46  & 18 & $1.7\times10^4$ & $1.1\times 10^6$ & 5 arm spiral in SW  US and Mexico \\
         Long Baseline & 30 & 18 & $1.3\times 10^5$ & $8.8\times 10^6$ & 10 stations of 3 antennas \\
         Short Baseline &  19 & 6 & 11 & 60 & close-packed pseudo-grid 
         \enddata
     
    \tablecomments{$^a$ Number of antennas. $^b$ Diameter of antennas. $^c$ Minimum baseline. $^d$ Maximum baseline. The Main array, used in this work, is the configuration where the Core + Spiral + Mid arrays are used simultaneously.}%
    \end{deluxetable*}

The design and construction of a new observatory need detailed preliminary studies to characterize and fine-tune its instrumental response, noise level, and resolving power. For radio interferometers, the point spread function (PSF) or the synthesized beam is directly related to the geographical distribution of the antennas. 
Therefore, investigating the effect of their location is an important part of the development of a new facility \citep[ngVLA memo \#92\footref{revD},][]{Trejo2024arXiv240702720T}. 
Additionally, by simulating observations, we can test the effects of the varying aggregated bandwidth in continuum observations, and of the spectral resolution in line observations, to have a realistic estimate of the noise and other imaging properties. This allows for the definition of the integration times that will be necessary for each type of astrophysical object and observing mode. 

In this paper, we generate synthetic observations to simulate and test the observational capabilities of the ngVLA.  
To achieve this, we used analytical models of the ionized gas distribution in two scenarios for the physical interpretation of young ionized regions around massive protostars: an ionized jet and an ionized disk.  
From these models, we perform post-processing radiative transfer to obtain surface brightness distributions, which serve as the ``sky model" input to generate interferometric data.
We perform ngVLA synthetic observations with varying observational parameters, such as channel width, total bandwidth, and integration time. 
With this analysis, we suggest the minimal observing parameters necessary to observe these regions once the observatory is operational. 

The organization of the paper is the following. In Section \ref{sec:methods} we describe
the procedure for generating synthetic ngVLA observations, including the use of \texttt{sf3dmodels} and \texttt{RADMC-3D} to generate analytical and radiative transfer models. 
In Section \ref{sec:jet_section} we describe the model and synthetic observations for an ionized jet. In Section \ref{disk_section} we describe the respective analysis for a gravitationally trapped ionized disk.
The ability of ngVLA to resolve these systems is discussed in Section \ref{sec:discussion}.
Our main conclusions are summarized in Section \ref{sec:conclusion}.

\section{Methods to Produce the ngVLA Synthetic Observations}\label{sec:methods}
We created models of ionized regions using the \texttt{sf3dmodels} package \citep{Izquierdo2018MNRAS.478.2505I}. This package allows for the production of 3D analytical models of star formation, ranging from isolated systems composed of disks, outflows, envelopes, and filaments, to complex systems that are the superposition of simpler ones \citep[e.g.,][]{Galvan-Madrid2018ApJ...868...39G, Carrasco-Gonzalez2021ApJ...914L...1C, Lin2022A&A...658A.128L}. Using this package, we generated the distributions of density, velocity, temperature, and ionization fraction. 

Next, we performed radiative transfer simulations of our analytical models using \texttt{RADMC-3D} \citep{Dullemond2012ascl.soft02015D}. 
In particular, we used the modified version that allows for the calculation of radio recombination lines \citep{Peters2012MNRAS.425.2352P}. This
produces data cubes where each channel contains a surface brightness map for a given frequency range (or channel width). These data cubes serve as ``sky models" representing the brightness distribution of the target object, which will be further processed to generate synthetic interferometric observations.

While single-dish telescopes directly measure the object's brightness distribution in the sky, radio interferometers like the ngVLA measure ``complex visibilities'', the Fourier components of the brightness distribution.  
Therefore, simulating interferometric observations first requires generating these visibilities. We use the \texttt{simobserve} task from the NRAO Common Astronomy Software Applications (\texttt{CASA}\footnote{\url{https://casa.nrao.edu}}) package \citep{CASA2022PASP..134k4501C} to create the visibilities, using our sky models as input. 
Key parameters include the position of the source, the central frequency of the observed bandwidth, the number of channels, the source integration time, the visibility integration time, and the antenna array configuration of the ngVLA.

This procedure produces a ``noiseless'' set of visibilities, to which we then add realistic noise consistent with the expected performance of the ngVLA. To do this, we first estimate the noise level for a naturally weighted image, $\sigma_{NA}$, using the \texttt{ngVLA Sensitivity Calculator}\footnote{\url{https://gitlab.nrao.edu/vrosero/ngvla-sensitivity-calculator}\label{ngVLA_calc}}. With this tool, once we define the instrument and weather-related noise, the total noise level depends mainly on the bandwidth and the time spent integrating on-source. 
We then calculate the noise per visibility, $\sigma_{simple}$,  with the following expression:
\begin{equation}
  \sigma_{simple} = \sigma_{NA} \sqrt{n_{ch} \, n_{pol} \, n_{baselines} \,  n_{integration}}, 
  \label{eq:noise_simple}
\end{equation}
where $n_{ch}$ is the number of channels and $n_{pol}$ is the number of polarizations (2 for Stokes $I$). The number of baselines is given by $n_{baselines} = N(N-1)/2$, with $N$ being the number of antennas, and $n_{integration}$ is defined by the time on-source divided by the integration time $t_{int}$, which is set to 10 seconds during this work.
Finally, the noise is added to the visibilities using the \texttt{CASA} simulation tools\footnote{\url{https://safe.nrao.edu/wiki/pub/ALMA/SimulatorCookbook/corruptguide.pdf}}. 
To transform the noisy visibilities into ``clean'' images we use the \texttt{tclean} task, which generates the synthetic ngVLA images.

\section{Ionized Jet Model} \label{sec:jet_section}
The first type of ionized region analyzed is a radio jet launched by a massive protostar \citep[e.g.,][]{Anglada2018,Purser2021}, constructed following the geometrical wind model of \textcite{Reynolds1986ApJ...304..713R}.  
This prescription has been widely used in the literature to model the thermal free-free emission of radio jets \citep[e.g.,][]{Torrelles97,Rodriguez05,Carrasco-Gonzalez2021ApJ...914L...1C,FeeneyJ23}.  
The model assumes that a jet with transverse radius $w_0$ is injected at a distance $r_0$ from the central object. Along the jet, the transverse radius depends on the distance to the injection point as $w(r) = w_0(r/r_0)^\epsilon$. The parameter $\epsilon$ defines the geometry of the jet, describing a collimated jet for values $\epsilon < 1$, and a poorly collimated wind for values of $\epsilon\ge1$. 
The physical properties of the gas in the jet, such as density, temperature, velocity and ionization fraction, also vary from $r/r_0$ to powers of $q_n$, $q_T$, $q_v$ and $q_x$, respectively. 

For the initial conditions, the jet is injected with a base number density of ionized gas $n_0$, temperature $T_0$, speed $v_0$, and ionization fraction $x_0$. Finally, mass conservation implies that the flow of material that crosses the area perpendicular to $r$, $A_\perp$, must be constant. This translates into the condition $n v A_\perp = \mathrm{constant}$, or $q_n= -q_v-2 \epsilon$.
If the jet is constructed with constant velocity ($q_v = 0$), the exponent of the density will be given by $q_n = -2 \epsilon$. Hence, the density at the injection radius will be given by the mass loss rate $\dot{M}$ as:
\begin{equation}
    n_0 = \frac{\dot{M}}{\pi \mu m_\mathrm{H} w_0^2 v_0}, 
\end{equation}
where $\mu$ is the mean particle weight and $m_\mathrm{H}$ is the mass of hydrogen atom. 

To test the capabilities of the ngVLA to observe and resolve this type of object, we model the ionized region as the sum of a collimated jet plus a poorly collimated wind, assuming the parameters estimated by \textcite{Carrasco-Gonzalez2021ApJ...914L...1C}
for the Cep A HW2 radio jet. This jet is driven by a protostar of mass $M = 15 ~ \mathrm{M_\odot}$ \citep[equivalent to an early B-type star, ][]{Anglada2018}. The collimated jet ($\epsilon_{\mathrm{jet}} = 2/3$) is injected at a distance $r_{\mathrm{0,jet}} = 25$ au from the central object, with a transverse radius $w_{\mathrm{0,jet}} = 5$ au, and constant velocity of $v_{\mathrm{0,jet}} = 500 \, \mathrm{km \, s^{-1}}$ \citep{Curiel2006}. The gas in the jet is completely ionized at all radii ($q_x = 0$) and isothermal ($q_T = 0$) with an electron temperature of $T_{\mathrm{0,jet}} = 10^4$ K. The mass-loss rate for this jet is $\dot{M}_{\mathrm{jet}} = 1.5 \times 10^{-6}$ $\mathrm{M_\odot}$  yr$^{-1}$.
The poorly collimated wind ($\epsilon = 1$) is injected at a distance $r_{\mathrm{0,wind}} = 3$ au from the protostar with a transverse radius $w_{\mathrm{0,wind}} = 5$ au. The ionization fraction at the injection radius is $x_0 = 10 \%$, with a power law index for the ionization fraction of $q_x = -0.5$.  
The gas in the wind has a constant velocity of $v_{\mathrm{0,wind}} = 100 \, \mathrm{km \, s^{-1}}$. The mass loss rate is set to $\dot{M}_{\mathrm{wind}} = 1.75 \times 10^{-6}$ $\mathrm{M_\odot}$  yr$^{-1}$.

\begin{figure}[!t]
    \centering
    \includegraphics[width = 0.75\textwidth]{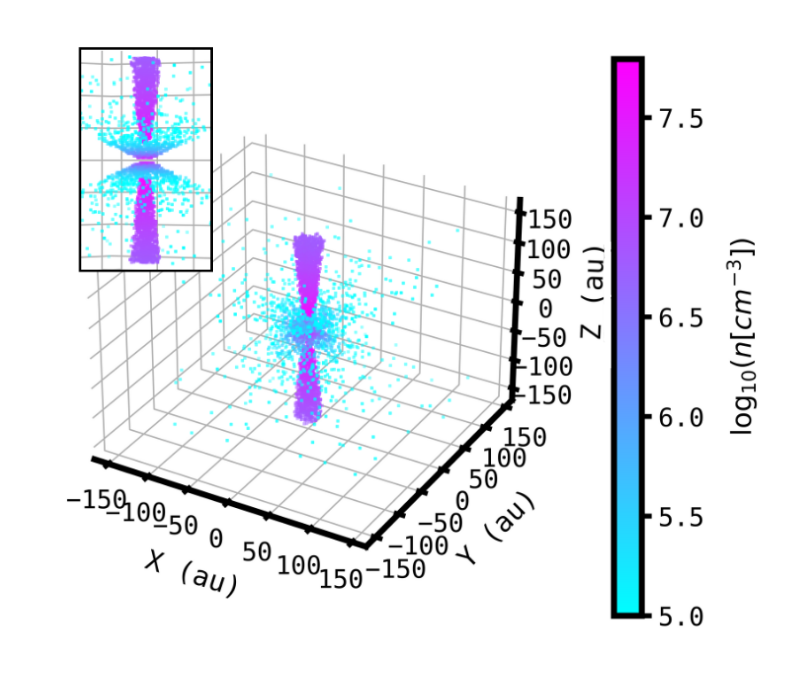}
    \caption{Ionized density distribution of the jet model. The main panel presents the model with an inclination of $60^\circ$ from the jet's axis ($Z$-axis). The smaller panel in the top left provides a view of the model from a line of sight perpendicular to the jet's axis. 
    The density of points and color scheme represent the ionized gas density.}
    \label{fig:density_model_jet}
\end{figure}

Figure \ref{fig:density_model_jet} shows the 3D density distribution of the ionized gas in this model. The ionized density at the model's center is dominated by the poorly collimated wind but decreases as we move away, and only the collimated jet can be seen at large distances from the protostar.

\subsection{Synthetic observation} 
\label{sec:synthobs_jet}
At higher frequencies the RRLs are brighter \citep{Peters2012MNRAS.425.2352P}, so we simulate observations with the ngVLA in Band 6 (central frequency 93 GHz) using the Main array, comprising the Core, Spiral, and Mid components, totaling 214 antennas. The antenna positions are specified in the ngVLA memo \#92\footref{revD}.
The minimum baseline for this configuration is $B_\mathrm{min} = 39$ m, while the maximum baseline is $B_\mathrm{max} = 1068$ km (see Table \ref{tab:Rev_D_configuration}). This range of baseline lengths traces angular scales from $\sim 17 \arcsec$ to $\sim 0.6$ mas at 93 GHz, respectively.\footnote{In reality the maximum recoverable scale is a factor $< 1\times$ of what corresponds to $B_\mathrm{min}$, depending on the probability density distribution of baseline lengths \citep[e.g.,][]{Ginsburg22}.}

To test the capability of the ngVLA to resolve this radio jet, we simulate two different kinds of observations, one for the H$41\alpha$ RRL ($\nu_0 = 92.034434
$ GHz) and another for the free-free continuum in Band 6 (tunable from 70 to 116 GHz). 
We configure the line observations with 80 spectral channels of 9.2 MHz ($30 \ \mathrm{km \, s^{-1}}$) each, centered at the line rest frequency. 
For the continuum observation, we use the total bandwidth (20 GHz) divided into 80 channels of 250 MHz each, centered at 93 GHz.  
The number of continuum channels is selected to have the coarsest possible channel width to reduce computation time, while simultaneously avoiding the effect of bandwidth smearing\footnote{\url{https://science.nrao.edu/facilities/vla/docs/manuals/oss2016A/performance/fov/bw-smearing}}.
For both cases, we use an on-source integration time of 5 hours.  The corresponding noise for a naturally-weighted image, $\sigma_{NA}$, is calculated using the \texttt{ngVLA Sensitivity Calculator}\footref{ngVLA_calc} for observations using the ngVLA Main array, with the bandwidth and observation times as defined above.
The central position of the jet in the sky corresponds to the phase center of the data set, which is set at (J2000) $\mathrm{R.A.} = 18^h 00^m 00^s.0$, $\mathrm{Dec} = +25^\circ 00' 00''.0$. The declination is chosen to simulate a source with good observability.

Using the setup described in Section \ref{sec:methods}, we generated visibilities using the \texttt{simobserve} task. Subsequently, noise per visibility, $\sigma_{simple}$, was calculated following Equation \ref{eq:noise_simple} and added with the \texttt{CASA} tools. 
Finally, clean images were produced using the \texttt{tclean} task in \texttt{CASA}.

We are interested in resolving the region where the jet is injected and collimated.
At the Cep A HW2 distance (700 pc) and inclination of $60^\circ$ of the jet's axis (Z-axis), the projected separations between the injection points of the collimated (50 au) and wide (6 au) components are 61.9 and 7.4 mas, respectively. Similarly, the transversal size at the base of the collimated and wide components is 10 au for both, or 12.4 mas. 

To achieve the desired balance between angular resolution and signal-to-noise ratio (S/N) to spatially and spectrally resolve the jet, we employed a combination of \emph{Briggs} weighting and $uv$ tapering within \texttt{CASA}'s \texttt{tclean} task. 
While this slightly broadens the synthesized beam compared to a uniform weighting, it enhances the S/N and improves the PSF shape (e.g., ngVLA memo \#55\footnote{\url{https://library.nrao.edu/public/memos/ngvla/NGVLA\_55.pdf}\label{memo55}} and memo \#89\footnote{\url{https://library.nrao.edu/public/memos/ngvla/NGVLA\_89.pdf}\label{memo89}}). To ensure a correct beam sampling, the images are 2400 pixels per dimension, with a pixel size of $0.35$ mas. This corresponds to a total field of view of $840$ mas.

For the continuum observation, the image was created using the multi-term multi-frequency synthesis (MTMFS) clean algorithm. We combine a robust parameter of $r = -0.5$ and a 2 mas $uv$ tapering to obtain an image with an rms noise of $0.41 \, \mathrm{\mu Jy \, beam^{-1}}$ and a synthesized beam of $2.5 \times 2.2 ~ \text{mas}^2$ with a position angle (P.A.) $= -24.9^\circ$.
For the line observation, we choose a robust parameter of $r = 0.0$ and a 1.25 mas $uv$ tapering, which results in an image with an rms noise of $12.4 \, \mathrm{\mu Jy \, beam^{-1}}$ and a median restoring beam of $9.6 \times 8.1\ \text{mas}^2$ and P.A. $=  89.5^\circ$.

The chosen weighting and tapering effectively reshape the PSF closer to a Gaussian profile, facilitating the deconvolution process during image cleaning. This is particularly advantageous for accurately recovering faint features while at the same time minimizing potential deconvolution artifacts. We also explored the use of uniform and natural weighting. On the one hand, uniform weighting resulted in a significantly lower S/N ratio, making reliable jet detection difficult. On the other hand, natural weighting results in a lower resolution compared to that needed for the scientific case presented here.
Currently, the ngVLA staff is working on characterizing the shape that the PSF of the ngVLA will have for different array configurations with different number of antennas, while varying the weighting, tapering, and declination parameters. 
Recently, \citet{Trejo2024arXiv240702720T} reported in ngVLA memo \#122\footnote{\url{https://library.nrao.edu/public/memos/ngvla/NGVLA_122.pdf}} on the positive effects of including antennas of the Main array in northern Mexico, both for the beam size and shape. 

\begin{figure}[th]
\includegraphics[width = \textwidth,trim={5cm 0 5cm 0},clip]{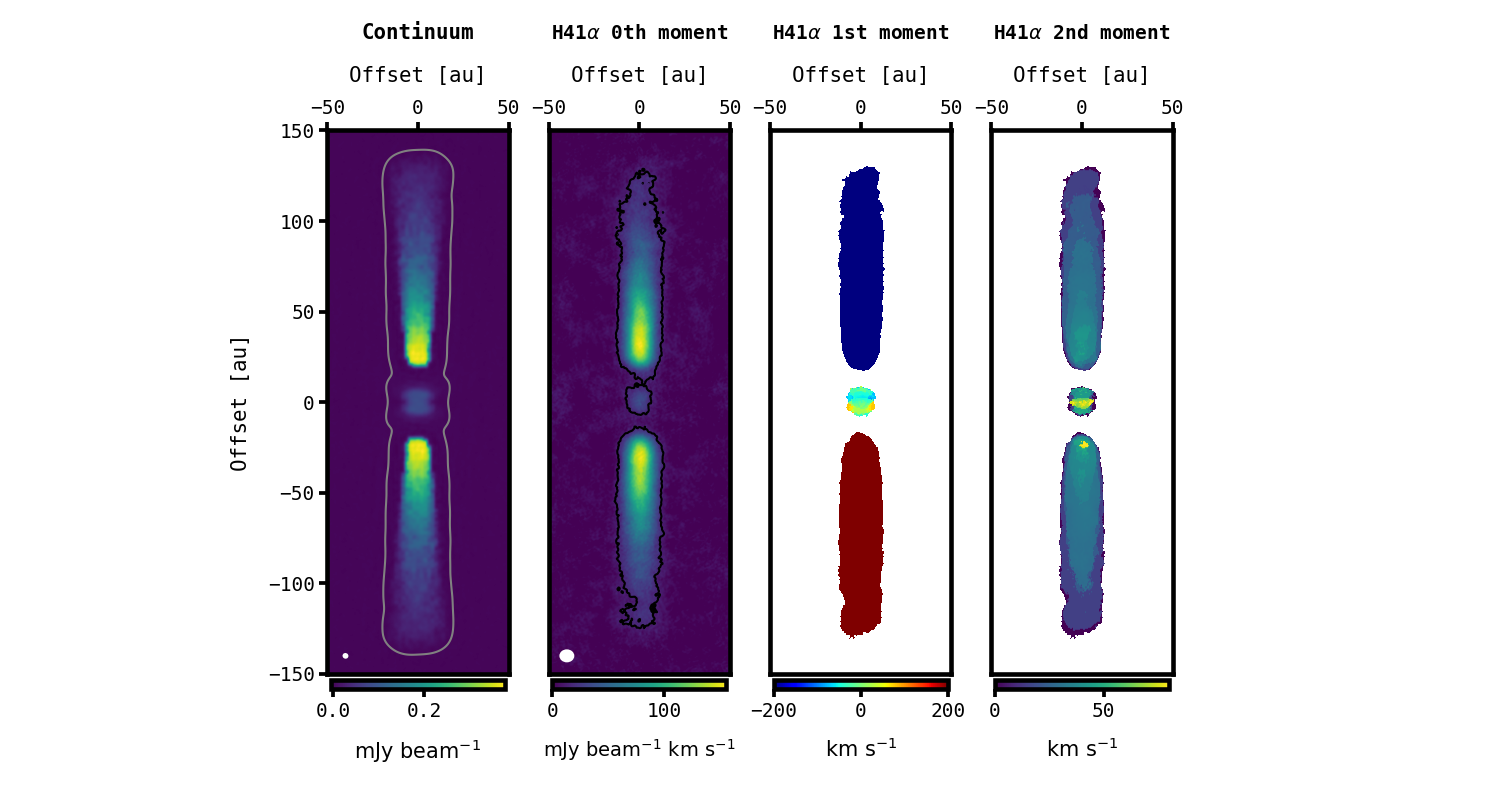}
    \caption{Synthetic observation of the jet model. The first panel, from left to right, shows the continuum emission map, where the gray contour mark the emission at $5\times$ the rms noise ($\sigma_\mathrm{cont,rms} = 0.41 \ \mathrm{\mu Jy \, beam^{-1}}$). The second, third, and fourth panels show the zeroth, first, and second moment maps of the H$41\alpha$ line, respectively. 
    In the second panel the black contours represent $10\times$ the zeroth-moment rms noise ($\sigma_\mathrm{0th,rms} = 2.9 ~ \mathrm{mJy \, beam^{-1} \, km \, s^{-1} }$). The first and second moment maps only used pixels in the cube with values larger than 5$\times$ the rms noise per channel ($\sigma_\mathrm{line,rms} = \mathrm{12.4 \, \mu Jy \, beam^{-1}}$) in the line observation.
    The synthesized beam ($\theta_\mathrm{FWHM} = 2.5  \times 2.2 ~\mathrm{mas}^2$) for continuum observations is shown in the lower left of the first panel.  The beam for the line observations ($\theta_\mathrm{FWHM} = 9.6 \times 8.1 ~\mathrm{mas}^2$) is shown in the lower left of the second panel.}
    \label{fig:jet_observation}
\end{figure}

To analyze the observations of the $\mathrm{H41\alpha}$ line emission we use moment maps. The zeroth moment is defined as the line emission integrated in velocity (or frequency) and is used to trace the total amount of line emission. 
The first moment is the intensity-weighted average velocity, traditionally used to obtain the ``velocity fields''. 
The second moment measures the velocity dispersion of the gas along the line of sight, i.e., the width of the spectral line.  
The second, third and fourth panels of Figure \ref{fig:jet_observation} show the moment maps for the simulated spectral cube. The zeroth moment map presents a morphology very similar to the continuum emission. This is expected as they both trace ionized gas.  
The first-moment map clearly shows the kinematics of the ionized gas, dominated by at least two velocity components: one high-velocity ($\lvert v \rvert \ge 200 \ \mathrm{km \ s^{-1}}$) component corresponding to the gas in the collimated jet; and one which is moving at lower velocities ($ \lvert v \rvert \leq 100 \ \mathrm{km \ s^{-1}}$), corresponding to the poorly collimated wind.

From the synthetic observation of the continuum emission, a 5-hour observation is sufficient to detect the complete extension of the collimated jet with a signal-to-noise S/N $> 50$, reaching a S/N $\sim 10^{3}$ at the peak of the emission. 
In the central part (20 au) we can detect the poorly collimated wind with a S/N $\sim 200$.
In the $\mathrm{H41\alpha}$ line observation, we detect the north and south parts of the jet with a $\mathrm{S_{peak}/N} \sim 70$. In the center, the wind component is detected with $\mathrm{S_{peak}/N} \sim 20$. 
Table \ref{tab:jet_obs_prop} summarizes the key properties of the ngVLA synthetic observations. 

\begin{deluxetable*}{cccccccc}
    \tabletypesize{\footnotesize} 
    \tablewidth{0pt}
    \tablecaption{Properties of ngVLA synthetic jet observations.  \label{tab:jet_obs_prop}}
    \tablehead{
        \colhead{Observation} & 
                \colhead{Number channels} &
        \colhead{Channel width} & 
        \colhead{Channel width} &
        \colhead{rms noise} & 
        \colhead{$S_\mathrm{peak}$} &
        \colhead{$T_\mathrm{b,peak}$} &
        \colhead{Beam} \\
        (Central frequency) &  &   \colhead{(MHz)}  & \colhead{($\mathrm{km \, s^{-1}}$)}  & \colhead{$(\mathrm{\mu Jy \, beam^{-1}})$} & 
         \colhead{$(\mathrm{mJy \, beam^{-1}})$}& 
         \colhead{(K)} & \colhead{($\mathrm{mas}^2$)}
         }
    \startdata
         Continuum (93 GHz)  & 80 &  250  & 805.9   & 0.41  
         &    0.39 
         & $9958\pm11$ & $2.5\times2.2$ \\
        Line (92.03 GHz)  & 80  &    9.2  & 30  & 12.4 & 0.99  & 
        $1858\pm23$ & $9.6\times8.1$ 
    \enddata
    \tablecomments{The resulting image noise from the \texttt{ngVLA sensitivity calculator} with natural weighting plus $uv$ tapering are $\sigma_\mathrm{cont, rms}=0.20 \, \mathrm{\mu Jy \, beam^{-1}}$ and $\sigma_\mathrm{line,rms}=10.5 \,  \mathrm{\mu Jy \, beam^{-1}}$ for the continuum and line, respectively.
    The noise reported for the continuum is for an observation using the entire available bandwidth; the noise in the line observations is for one channel. The integration time is 5 hours. $T_\mathrm{b,peak}$ is the peak brightness temperature.}
\end{deluxetable*}

While in the work of \citet{Carrasco-Gonzalez2021ApJ...914L...1C} the continuum emission of the radio jet in Cep A HW2 is partially resolved with a beam size of $42\times28$ mas$^2$ using the VLA, in this work we demonstrate that the ngVLA will be able to resolve this region with $>10\times$  better angular resolution and significantly better sensitivity.
Moreover, unlike in our model, real
jets present substructures such as shocks and knots that the ngVLA will be able to resolve.

\begin{figure}
    \centering
\includegraphics[width=\textwidth]{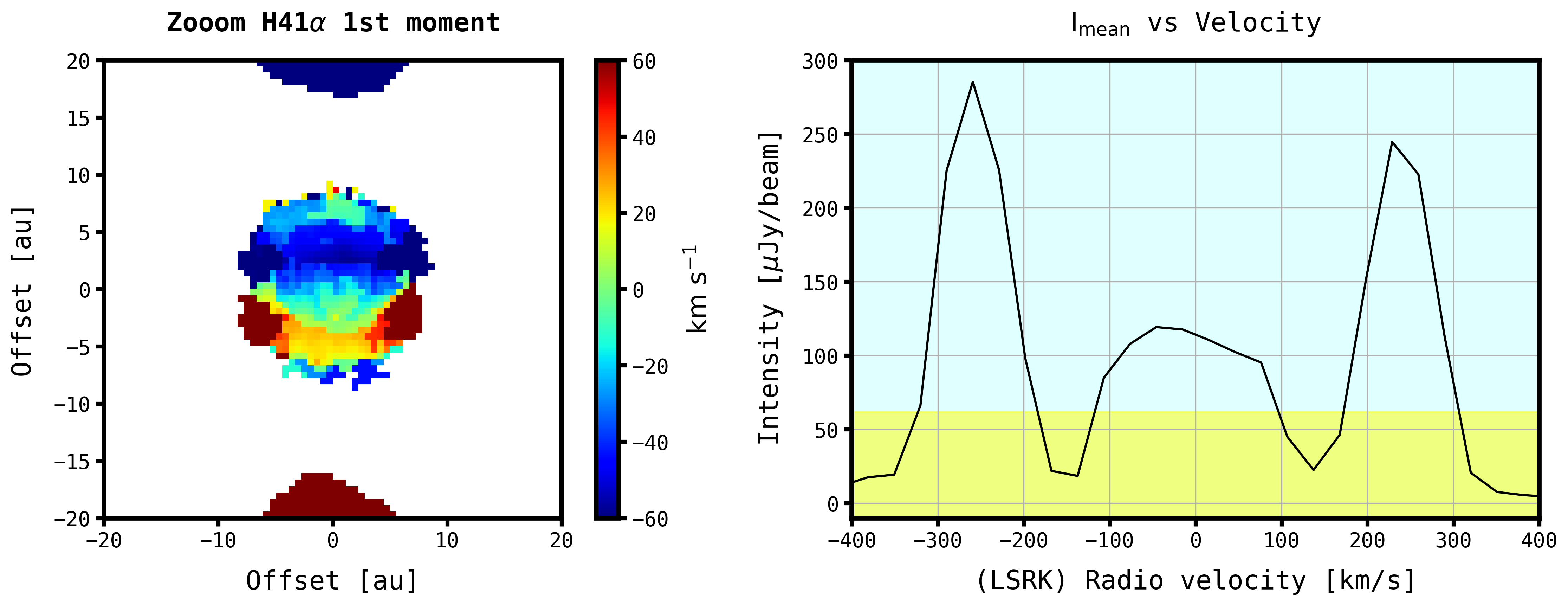}
\caption{The left panel shows a zoom-in toward the central part of the jet+wind model of the first moment of the $\mathrm{H41\alpha}$ line. The right panel shows the spectrum averaged over a region enclosing the jet and poorly collimated wind. The yellow shaded area represents emission below five times the average rms noise.}   \label{fig:central_zone_moment1}
\end{figure}

In addition to the continuum, the ngVLA will be able to resolve for the first time the emission from radio recombination lines\footnote{A tentative detection of millimeter recombination lines in the jet of Cep A HW2 was reported by \citet{JimenezSerra2011}.}, allowing us to characterize the resolved kinematics of this type of object.
As the first moment map shows (third panel, Figure \ref{fig:jet_observation}), the ngVLA can resolve the distinct kinematical components: the blue- and red-shifted parts of the collimated jet and the respective sides from the poorly collimated wind, mostly close to the center. 
A zoom-in on the central area of the first moment map (first panel, Figure \ref{fig:central_zone_moment1}), shows the velocity fields of the different components of the ionized gas. 
When we analyze the spectrum (second panel Figure \ref{fig:central_zone_moment1}), we can also notice that it shows the two kinematic components: the slow wind at lower and the fast jet at higher velocities.

In summary, ngVLA observations of radio jets will be able to differentiate between the existence (or coexistence) of collimated jets and wide-angle winds and resolve their launching radii, widths, and any possible substructure down to a few au, both in continuum and in recombination line emission.

\section{Ionized disk}
\label{disk_section}
The ionized disk model is motivated by the expectation that nuclear fusion begins before the accretion phase of high-mass star formation is complete. 
This generates a large Lyman photon flux ($E>13.6$ eV) from the protostar, which ionizes the neutral material around it, i.e., it creates an \textsc{H\,ii} region that eventually clears out the parent molecular cloud \citep[e.g.,][]{Arthur2011, Jaquez23}. 
However, it is unclear to what extent accretion produces a ``bloating'' of the (proto)stellar surface and delays the first appearance of these \textsc{H\,ii} regions \citep{HYO2010}.
If the very young HC \textsc{H\,ii} region is approximately smaller than its ``gravitational radius'' $R_g =  G M / 2 c_s^2$, where $c_s$ is the sound speed of the ionized gas, the \textsc{H\,ii} region will be confined by gravity  \citep{Keto2003ApJ...599.1196K,Keto2007ApJ...666..976K,Peters2010ApJ...719..831P}. 
It is within this framework that we test the performance of the ngVLA to image ionized accretion disks and envelopes around massive (proto)stars. 

For our model, we assume a Keplerian, rotationally supported, flared disk truncated at a radius $R_d$.
We use the standard \textcite{Pringle1981ARA&A..19..137P} disk model, for which the density distribution is given by:
\begin{equation}
    n (R,z) = n(R) \exp [ -z^2 /[2H(R)^2]],
\end{equation}
where $R$ and $z$ are the cylindrical coordinates for the polar radius and height. 
$H(R)$ is the scale height defined as:
\begin{equation}
    H(R) = H_0 (R/R_d)^q,  
    \label{eq:height_disk_equation}
\end{equation}
where $H_0$ is a normalization constant chosen so that $H(R_d) = R_g$, i.e., the ionized disk is thick. We have also added a condition that only the gas within a sphere of radius $R_g$ is ionized, thus our model does not represent any unbound ionized gas at larger radii. 

The midplane radial density profile $n(R)$ is defined as:  
\begin{equation}
    n(R) = n_{d0} (R/R_d)^p, 
    \label{eq:density_disk_equation}
\end{equation}
where $n_{d0}$ is the normalization at $R_d$. 
To define the density of the mid-plane of the disk, we use the model of \textcite{Keto2010MNRAS.406..102K} for non-spherical accretion, where we assume that the ionized disk connects with an envelope of neutral material at a radius $R_d$. Those authors modeled the envelope following the prescription presented by \textcite{Mendoza2004RMxAA..40..147M}, which is a modified version of the original \textcite{Ulrich_1976ApJ...210..377U} model.  
Following this approach, we can relate the mass accretion rate, $\dot{M}$, and the midplane  density,  $n_{d0}$, as 
\begin{equation}
    \dot{M} = 4 \pi  R_d^2 n_{d0}  v_k,
\end{equation}
where $v_k = \sqrt{GM/R_d}$ is the Keplerian velocity at radius $R_d$. 

In this approach, the model is primarily defined by $R_g$, which is given by the mass of the central protostar, and the mass accretion rate $\dot{M}$ of the envelope.
We assume a central protostar $M = 20 \, \mathrm{M_\odot}$, giving $R_g = R_d = 119.5$ au. 
To establish the mass accretion rate, we suppose that the protostar emits a flux of hydrogen-ionizing photons similar to that of a main-sequence star, with $\log Q_\mathrm{Ly} = 48.2$ s$^{-1}$ \citep{Martins2005A&A...436.1049M}. To maintain ionization in the disk, we equate the total number of recombinations with the ionizing photon flux:
\begin{equation}
    Q_\mathrm{Ly} = \alpha_B \sum_i dV_i n_{i}^{2},
\end{equation}
where $dV_i$ and $n_{i}$ represent the volume and density of particles in the $i$-th cell within the ionized disk, respectively. Here, we use $\alpha_B =2.6\times10^{-19} \, \mathrm{m^{3} \, s^{-1}}$, the case B recombination coefficient. Based on these assumptions, we set the mass accretion rate to $\dot{M} = 3.2 \times 10^{-5} \, \mathrm{M_\odot \, yr^{-1}}$. 

\begin{figure}[tbh]
    \centering
    \includegraphics[width = 0.75 \textwidth]{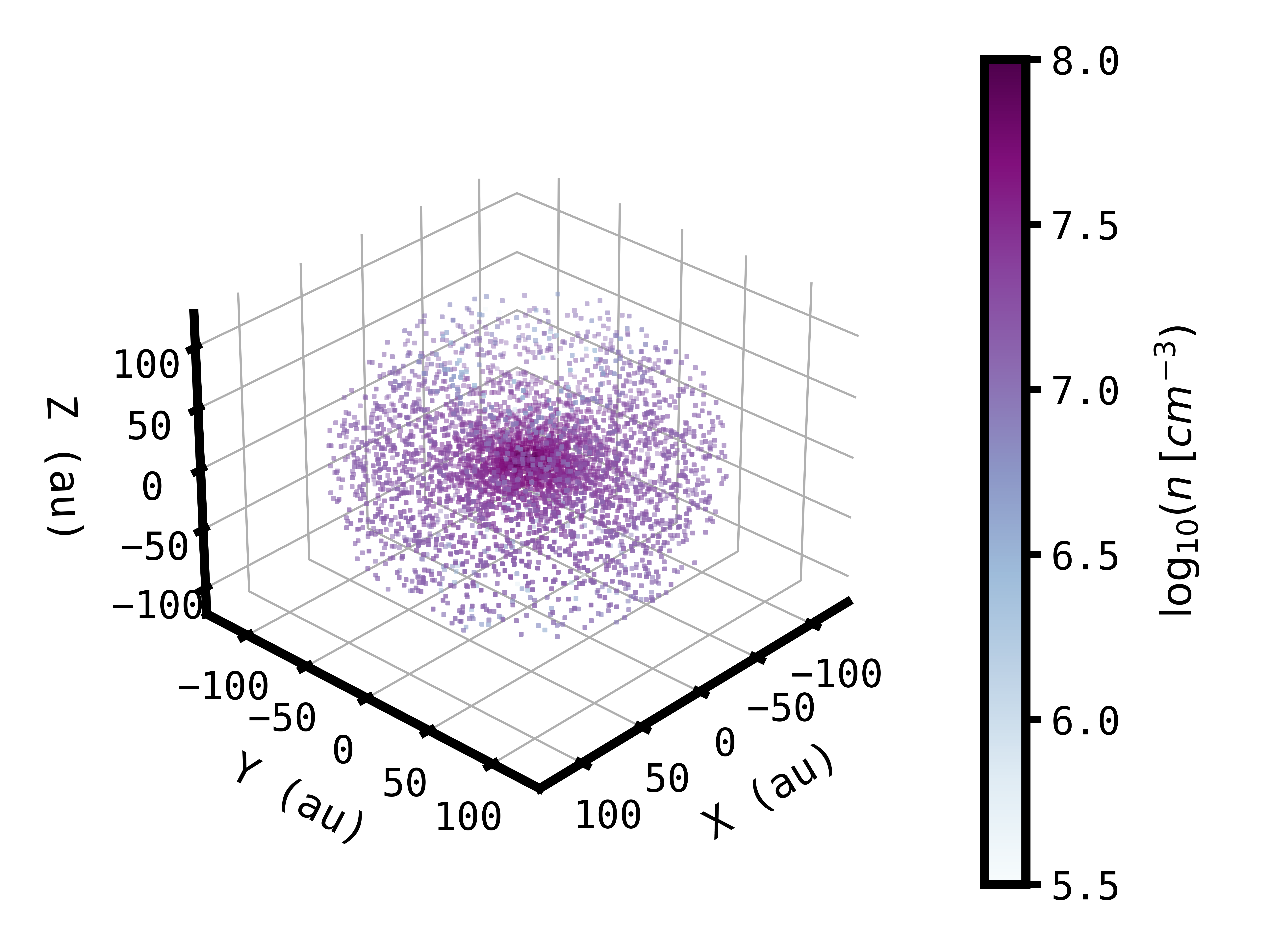}
    \caption{Ionized density distribution of the disk model. The density of points and color scheme represent the ionized gas density.}
    \label{fig:disk_3d_plot}
\end{figure}

We assume that the ionized disk surrounding a single massive star will behave similarly to the model that has been applied to the dense stellar cluster at the center of G10.6 \citep{KW2006}. The power-law indices for the scale height in Equation \ref{eq:height_disk_equation} and in the density Equation \ref{eq:density_disk_equation} are set to $q = -1$ and $p = -0.674$, following the best-fit model of the gravitationally trapped HC \textsc{H\,ii} region in G10.6 by \citet{Galvan-Madrid2023ApJ...942L...7G}.
Finally, we set the temperature of the ionized gas in the disk to $T_{\mathrm{e}} = 9 \times 10^3 $ K \citep{Osterbrock2006agna.book.....O, Wilson2013tra..book.....W}.
Figure \ref{fig:disk_3d_plot} shows the 3D ionized density distribution of the disk model. 

\subsection{Synthetic observation} \label{sec:synthobs_disk}
Our primary objective with this model is to kinematically resolve the ionized disk. In particular, to resolve the characteristic Keplerian profile, i.e., higher velocities closer to the star, that manifests in typical position-velocity (PV) diagrams of rotating disks \citep[e.g.,][]{Johnston2015ApJ...813L..19J,Maud2018,Ginsburg2018,Zapata2019ApJ...872..176Z,Ahmadi2023A&A...677A.171A}. 

For this model, we simulate one observation for the H38$\alpha$ RRL ($\nu_0 = 115.274399$ GHz) and another for the ngVLA Band 6 continuum emission.
We configure the line observation with 60 channels of 1.9 MHz ($5 ~\mathrm{km \,  s^{-1}}$) each, centered at the line rest frequency. 
For the continuum observation, we divide the total continuum bandwidth of 20 GHz into 60 channels of 333 MHz each, centered at 106 GHz. The integration time on-source is also set to 5 hours in both cases. We place the disk at a distance of 4 kpc with an inclination of $30^\circ$ with respect to the rotation axis (Z-axis).
\begin{deluxetable*}{c c c c c c c c}
    \tabletypesize{\footnotesize}
    \tablewidth{0pt}
    \tablecaption{Properties of the ngVLA synthetic disk observations.  \label{tab:disk_obs_prop}}
    \tablehead{
        \colhead{Observation} & 
        \colhead{Number channels} &
        \colhead{Channel width} & \colhead{Channel width} & 
        \colhead{rms noise}      & 
        \colhead{$S_\mathrm{peak}$} & 
        \colhead{$T_\mathrm{b,peak}$} &
        \colhead{Beam} \\
        (Central frequency) &  &    (MHz)  & ($\mathrm{km \, s^{-1}}$) & $(\mathrm{\mu Jy\, beam^{-1}})$& $(\mathrm{mJy \, beam^{-1}})$ & (K) & ($\mathrm{mas^2}$)}
    \startdata
     \hline
        Continuum (106 GHz) & 60  & 333 & 867 &    0.46    &   0.44 & $9158\pm10$  &  $2.4\times2.2$ \\
        Line (115.3 GHz) &  60 & 1.9  & 5  &   117  &  10 
        & $55690\pm230$ & $4.8\times4.3$ 
    \enddata
    \tablecomments{The resulting image noise from the \texttt{ngVLA sensitivity calculator} with natural weighting plus $uv$ tapering are $\sigma_\mathrm{cont, rms}=0.28 \, \mathrm{\mu Jy \, beam^{-1}}$ and $\sigma_\mathrm{line,rms}=63 \,  \mathrm{\mu Jy \, beam^{-1}}$, for the continuum and line, respectively.
    The noise reported for the continuum is for an observation using the entire available bandwidth; the noise in the line observations is for one channel. The integration time is 5 hours. $T_\mathrm{b,peak}$ is the peak brightness temperature.}
\end{deluxetable*}

\begin{figure}[th]
    \includegraphics[width = \textwidth]{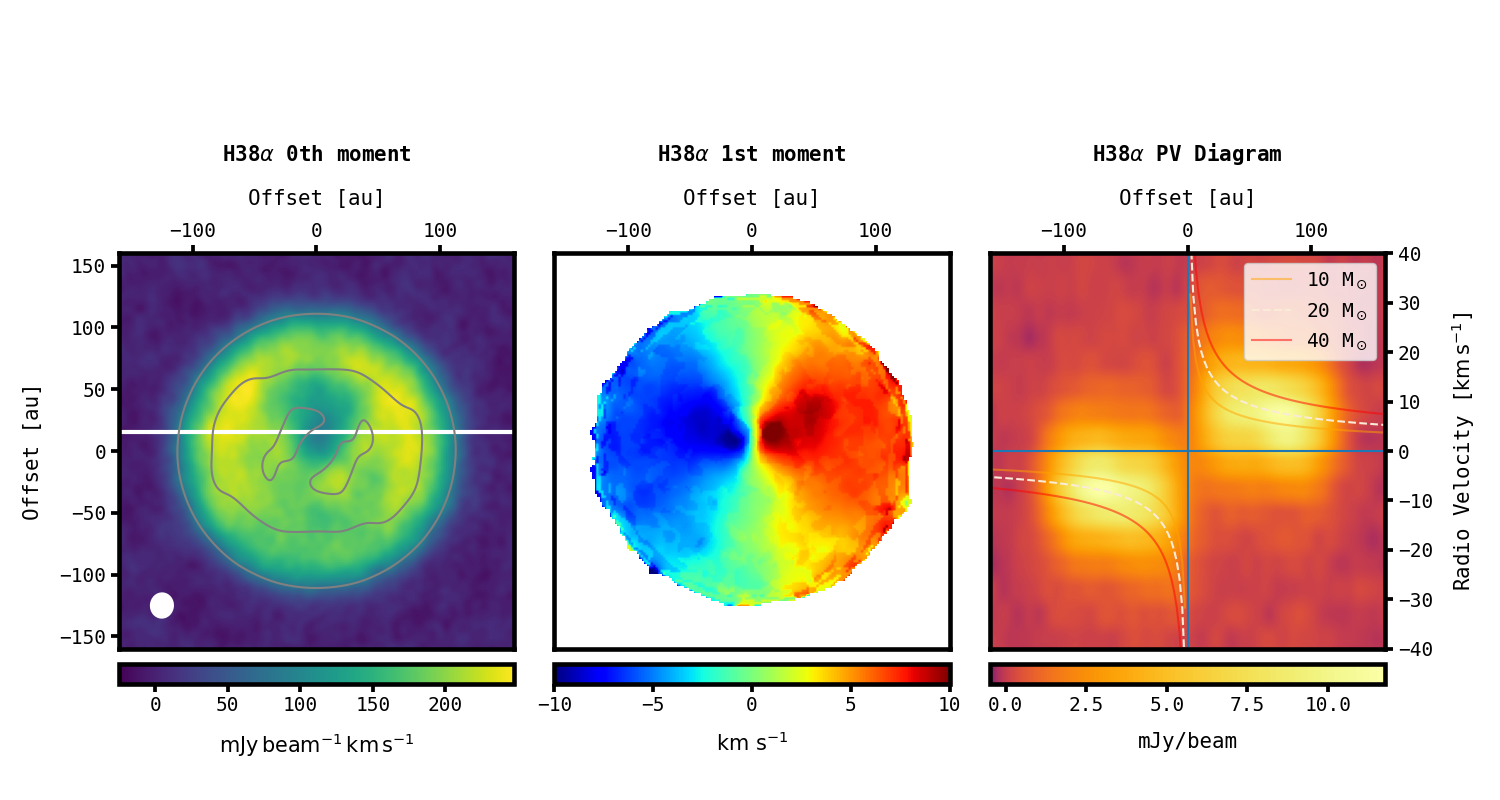}
\caption{Synthetic observation of the disk model. The first panel, from left to right, shows the zeroth moment map, where the gray contours represent the continuum brightness temperature at 5000, 8500, and 9000 K.
The second panel shows the first moment map. The third panel shows the PV diagram along the white line in the first panel, where the orange, dashed white, and red lines draw the rotation curves for central masses of 10, 20, and 40 $\mathrm{M_\odot}$. The synthesized beam ($\theta_\mathrm{FWHM} = 4.8 \ \times 4.3 \ \mathrm{mas}^2$) for the line observation is shown in the lower left corner of the first panel.
\label{fig:synthetic_observations_disk}}
\end{figure}

The visibilities are generated using \texttt{simobserve} with the previously described setup. Subsequently, the noise per visibility was added. Finally, we produce the image using the \texttt{tclean} task in \texttt{CASA}. 
As discussed in Section \ref{sec:jet_section}, using a combination of Briggs weighting and $uv$ tapper provides an advantage in reconstructing the PSF for the final images. For this model we use a robust value of $r=-0.5$ and a $uv$ taper of $2$ mas, yielding synthesized beams with an FWHM of $2.4 ~ \times 2.2 ~ \mathrm{mas}^2$ for the continuum and FWHM = $4.8 ~ \times 4.3 ~ \mathrm{mas}^2$ for the line observation.\footnote{It is known that \texttt{CASA} tends to produce larger beams for cube deconvolution than for continuum images of the same data set \citep[see, e.g., Section 3 in][]{Cunningham23}. The main reason likely is the improved $uv$ coverage when using the multi-term multi-frequency synthesis \citep[MTMFS,][]{RauCornwell2011} algorithm during continuum imaging .}  
This allows to resolve the emission across the disk.
The final images are 2400 pixels per spatial dimension, with a cell size of $0.25$ mas (field of view of $600$ mas).
Table \ref{tab:disk_obs_prop} summarizes the main properties of the generated ngVLA synthetic observations and Figure \ref{fig:synthetic_observations_disk} shows the resulting moment maps and the PV diagram. 

Interestingly, we found that the line peak brightness temperature in the disk model presents values much greater than those expected for local thermodynamic equilibrium (LTE, $T_B \leq 9\times10^{3} \ \mathrm{K}$), which means that the recombination line must be a maser \citep[e.g.,][]{JimenezSerra2013, Prasad2023}.  
From the disk model grid (see Figure \ref{fig:disk_3d_plot}), it turns out that a good fraction of the model volume has densities $n \sim 10^7 \ \mathrm{cm^{-3}}$. Under these conditions, the H$38\alpha$ line can be effectively amplified \citep{Walmsley1990}.

\begin{figure}[bt]
    \centering
    \includegraphics[width=\textwidth]{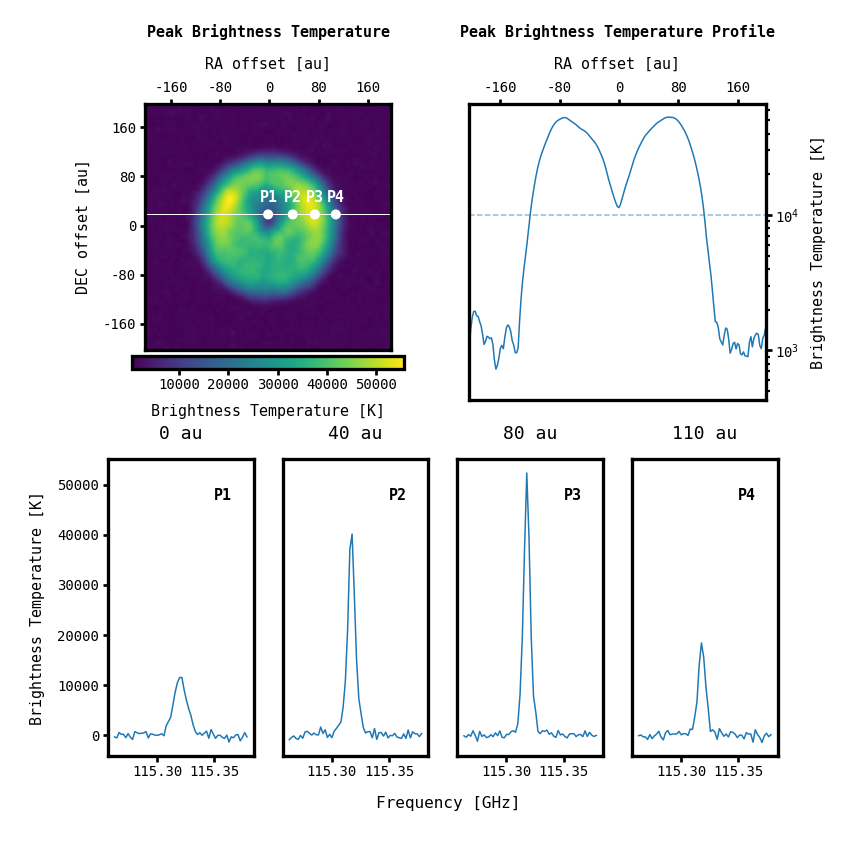}
    \caption{The upper left panel shows a map of the maximum value of the spectrum of the H$38\alpha$ data cube (moment-8 map). The upper right panel shows the peak brightness temperature profile along the white line in the upper left panel.  The bottom panels show the H$38\alpha$ line spectra for the pixels centered in the white circles in the upper left panel labeled as P1, P2, P3, P4.}
    \label{fig:intensity_cut}
\end{figure}

To visualize the spatial distribution of brightness temperature and determine the radii at which the line is amplified, we use the so-called moment-8 map, which shows the maximum value of the spectrum at each position (top-left panel of Fig. \ref{fig:intensity_cut}) and a horizontal cut across this map (top-right panel). 
This cut shows a spatial profile of peak intensities regardless of the channel where they are located. 
We find that the maser emission reaches its maximum intensity at a radius of approximately 80 au. The profile also reveals that in the central region of the disk, where the density is higher, the line emission is not amplified. This maser ``quenching'' at higher densities causes the brightness temperature in the disk center to return to values closer to LTE.
Furthermore, the spectra from pixels at different radii (bottom panels of Figure \ref{fig:intensity_cut}) also show that the line peak brightness is highest around a radius of 80 au, rapidly decreasing away from it. 

The kinematics of the disk is shown in the center (first moment) and right (PV diagram) panels of Figure \ref{fig:synthetic_observations_disk}. As expected for a Keplerian disk, we observe lower velocities at the outer regions and increasingly higher velocities closer to the center. 
The PV diagram shows that ngVLA observations can be used to retrieve the original input mass of the protostar responsible for ionizing the accretion disk. 
The figure shows the PV loci for masses of 10, 20,  and 40 $\mathrm{M_\odot}$.

\section{Discussion}  \label{sec:discussion}
\subsection{Understanding early ionization in massive star formation}
In this paper, we used analytical models of ionized regions (a jet and a gravitationally-trapped disk) to create ngVLA synthetic observations. These simulations defined the angular resolution and sensitivity needed to resolve the relevant free-free continuum and recombination line emission. 
Given the physical properties (mass, velocity, distance) of the models, we can define the situations that the ngVLA will be able to detect and resolve. This type of simulation can guide the setup and processing of real ngVLA observations, including integration time, bandwidth, number of channels, and deconvolution parameters. 

The ngVLA, like its predecessors the VLA and ALMA, will begin a new era in the study of the cosmos.
Regarding the study of the youngest \textsc{H\,ii} regions and radio jets, it will greatly expand our observational sample.
By carrying detailed studies of young stellar objects that contain ionized regions, we will be able to characterize the physical processes responsible for the earliest feedback from massive protostars \citep[e.g.,][]{Rosero2016ApJS..227...25R,Maud2018,ZhangY2019a,ZhangY2019b,Rosero2019ApJ...880...99R}. 
Those new observations will provide important clues to our understanding of how stars are assembled, mainly those of high mass.
Likewise, we will be able to understand the interaction of this ionized gas with the molecular envelope and the surrounding interstellar medium in general, improving our understanding of the feedback processes that halt accretion and set the final stellar masses. 

\subsection{Radio jets in the ngVLA era}
In section \ref{sec:jet_section} we showed that the ngVLA will fully resolve bipolar radio jets at distances of $700$ pc, like the case of Cep A HW2, in observations of only a few hours.
Therefore, objects in nearby (100 to 500 pc) star-forming regions will be studied in great detail. In Appendix \ref{appendix} we explore the limits of the ngVLA's capabilities. Using simple synthetic observations, we determined that the base of a radio jet similar to Cep A HW2 can be resolved up to distances of 2 kpc. 
Additionally, we find that the faintest radio jet that ngVLA can resolve with the selected observing parameters and at a distance of 700 pc is associated with a central object of a few solar masses. Therefore, the ngVLA will be able to resolve radio jets from lower-mass protostars in the nearby clouds located at distances of $\sim 100$ to 500 pc.

The resolution and sensitivity of the ngVLA will allow us not only to separate the components that make up the radio jet, but also to resolve their morphology and kinematics.
By resolving the jet both along the axis of the jet and in the direction perpendicular to it, we will be able to distinguish between external \citep[e.g.,][]{Albertazzi2014} or internal \citep[e.g.,][]{Shang2023} collimation scenarios \citep[e.g.,][]{Carrasco-Gonzalez2021ApJ...914L...1C,RK2022,FeeneyJ23}. 
With this, we will be able to improve our models and understand how jets disperse the angular momentum of the accretion flow \citep[e.g.,][]{Zapata2015} and how they interact with the surrounding ISM \citep[e.g.,][]{RK2019}.

\subsection{Ionized disks in the ngVLA era}
With the synthetic observations presented in Section \ref{disk_section}, we demonstrated that the ngVLA will be able to fully resolve, for the first time, the ionized disks around individual protostars \citep[e.g.,][]{Guzman14,Maud2018}. Observations of radio recombination lines will allow us to describe the morphology and kinematics, and rotation curves will serve to constrain the masses of the central objects through radiative transfer modelling \citep[e.g.,][]{Galvan-Madrid2023ApJ...942L...7G}.

Furthermore, our model predicts that the ionized regions surrounding a protostar with a mass of $20 \ M_\odot$  will exhibit maser amplification of the H$38\alpha$ line. This indicates that the ionized regions around such massive protostars can produce unusually bright recombination lines, which may enhance our ability to detect and study these regions \citep[e.g.,][]{JimenezSerra2013,Prasad2023}. 
In Appendix \ref{appendix}, we demonstrate that ngVLA can resolve an ionized disk produced by a $20~\mathrm{M_\odot}$ protostar located at a distance of 12 kpc, i.e., even further away than the Galactic center. 
Additionally, we find that the ionized disk produced by a star of only $10~\mathrm{M_\odot}$, the lowest mass star capable of producing an \textsc{Hii} region, can be resolved at a 4 kpc distance.

This opens the door to the study of the latest stages of accretion in individual protostars with masses greater than $20~ \mathrm{M_\odot}$ in more distant regions \citep[e.g.,][]{ZhangY2019b}. In addition, we will be able to observe ionized disks around massive protostars in nearby star-forming regions.
Thus, the ngVLA will allow us to investigate the conditions of massive star formation throughout the Galaxy, expanding our understanding of star formation at different mass scales and environments, helping to understand the mechanisms that cause stars to limit their final mass.

\section{Summary and conclusions} \label{sec:conclusion}
Studying the physical processes around massive protostars, such as jets, outflows, and disks, is crucial for understanding how accretion proceeds and stops, and how the final mass of stars is set. 
At some point in their early lives, massive protostars start hydrogen burning while material continues to fall from their surrounding envelopes, giving rise to the earliest \textsc{H\,ii} regions. Since these protostellar objects remain embedded in their natal clouds, they are better studied at radio and millimeter wavelengths. Characterizing them is key to understanding the formation of massive stars and their clusters. 

Currently, the VLA is only capable of detecting and resolving the radio continuum of nearby radio jets, or the continuum and recombination line emission of a few peculiar ionized disks.  With the future ngVLA, it is expected that astronomers will be able to routinely resolve the continuum and kinematics of individual massive protostars at the typical kiloparsec distances of massive star formation regions.

This work explores the capabilities of the ngVLA to resolve the continuum and recombination line emission from HC \textsc{H\,ii} region precursors. 
Two distinct analytical models of these types of ionized regions were used: a radio jet and a gravitationally trapped ionized disk. Radiative transfer simulations were performed to obtain their surface brightness distributions, which served as sky brightness models to create synthetic observations with the ngVLA instrument response.
We find that ngVLA observations can spatially and spectrally resolve these two types of ionized regions. With integration times of approximately 5 hours, radio jets driven by $10 \ \mathrm{M_\odot}$ protostars can be well resolved at distances of 700 pc. Additionally, ionized disks around $20 \ \mathrm{M_\odot}$ stars can be resolved at distances of 4 kpc.

\begin{acknowledgments}
This work was supported by the ngVLA Community Studies program, coordinated by the National Radio Astronomy Observatory (NRAO), which is a facility of the National Science Foundation (NSF) operated under cooperative agreement by Associated Universities, Inc.
This work benefited from the UNAM-NRAO Memorandum of Understanding in the framework of the ngVLA Project (MOU-UNAM-NRAO-2023).
J.J.-D. would like to thank CONAHCyT scholarship No. 760668. 
J.J.-D., R.G.-M., C.C.-G., and A.P. acknowledge support from CONAHCyT Ciencia de Frontera project ID 86372 ``Citlalcóatl''. 
R.G.-M. and J.J.-D. also thank the support from UNAM-PAPIIT project IN108822 and IN105225. 
C.C.-G. acknowledges support from UNAM DGAPA-PAPIIT grant IG101224.
L.A.Z. acknowledges financial support from CONAHCyT-280775, UNAM-PAPIIT IN110618, and IN112323 grants, México.
\end{acknowledgments}

\software{Astropy \citep{astropy:2018,astropy:2022}, 
\texttt{CASA} \citep{CASA2022PASP..134k4501C},
\texttt{RADMC-3D} \citep{Dullemond2012ascl.soft02015D},
\texttt{sf3dmodels} \citep{Izquierdo2018MNRAS.478.2505I}.}

\appendix
\section{Observational limits} 
\label{appendix}
We evaluate the capability of the ngVLA to detect regions that are both more distant and intrinsically fainter than those analyzed in Sections \ref{sec:jet_section} and \ref{disk_section}.
To achieve this, we create simplified synthetic observations of radio jets and ionized disks, including the effects of resolution and noise, but not including the full interferometric response. We consider one model with more restrictive physical parameters (i.e., a model that is intrinsically less luminous) and another model located at a further distance. These observations use the same number of channels and bandwidth as outlined in Sections \ref{sec:synthobs_jet} and \ref{sec:synthobs_disk} for the continuum and line observations of both the jet and disk models.

\begin{figure}
    \centering
    \includegraphics[width=0.65\linewidth]{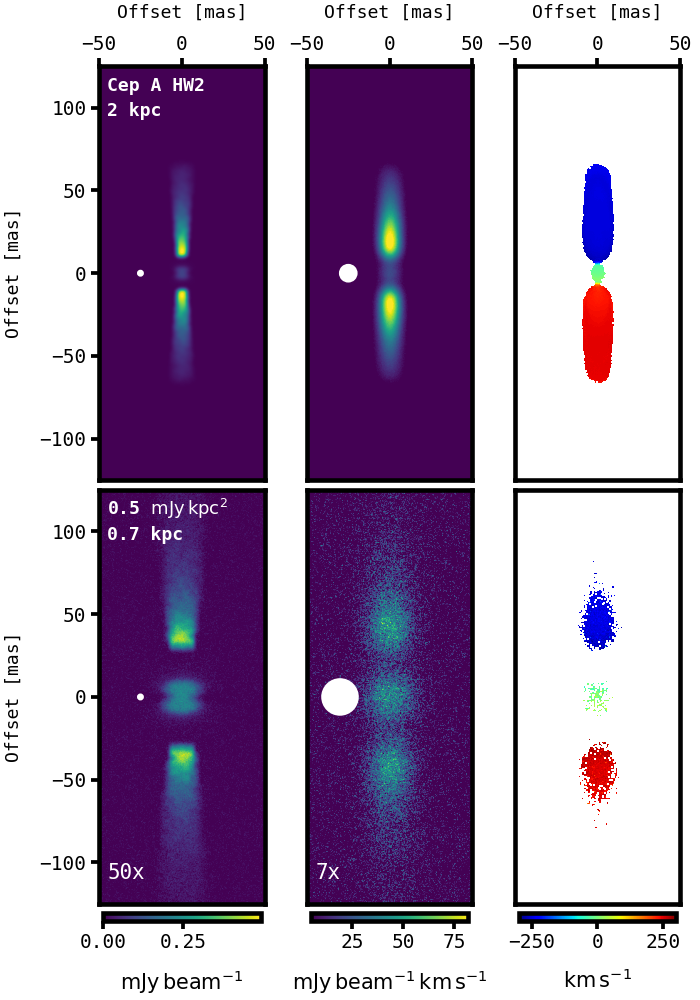}
    \caption{Simple synthetic observations of additional jet models.
    The columns show, from left to right, the continuum, zeroth, and first moment maps. 
    The top row shows the CepA HW2 model located at a further distance of 2 kpc. The second row shows the model of  a jet with radio luminosity $=0.5 \ \mathrm{mJy \, kpc^{2}}$ at 5.8 GHz, located at 700 pc. Note that the continuum and zeroth moment images do not have the same color scaling. The white number in the first and second panel of the second row is the scaling factor applied to the displayed intensity.}
    \label{fig:jet_param_models}
\end{figure}

For the Cep A HW2 radio jet, we determined the maximum distance at which it can be mapped in well resolved observations, based on achieving at least six beams between the bases of the two parts of the collimated jet in continuum observations. Setting the source at 2 kpc, the resulting simplified synthetic observations are shown in the top row of Figure \ref{fig:jet_param_models}.
At this distance, the model radio jet remains detectable and well-resolved in continuum observations, with a high S/N for both the jet and the central portion of the poorly collimated wind. The north and south parts of the collimated jet are still clearly resolved, enabling us to determine the injection distance from the protostar using both continuum and line observations.

Additionally, we model an intrinsically fainter jet with a radio luminosity of  0.5 $\mathrm{mJy \, kpc^{2}}$ at 5.8 GHz, which is three times less luminous than the Cep A HW2 model. This jet corresponds to a YSO with bolometric luminosity $L \sim 800~L_\odot$ \citep{Purser2021}. If the central star is already in the main sequence and using a luminosity-to-mass relation $L \propto M^{3.5}$ and the stellar calibration of \citet{Martins2005A&A...436.1049M}, this corresponds to a stellar mass of about $5~\mathrm{M_\odot}$. The mass of the central YSO would be smaller if it is still powered by accretion. 
This fainter jet is placed at a distance of 700 pc. For this model, the bases of the collimated component and the poorly collimated wind are still detected with S/N $>5$ in continuum observations. However, the line emission is too weak to be detected with the current setup. To improve the S/N, we use a beam twice the size (FWHM $\sim 21$ mas) of that used in the Cep A HW2 model. This allows us to detect the line emission with a S/N $>3$ The resulting synthetic observations are shown in the bottom row of Figure \ref{fig:jet_param_models}.

For the ionized disk, we analyze the model presented in Section \ref{disk_section}, which has a gravitational radius of $\sim 120$ au and a diameter of 240 au. When placed at a distance of 12 kpc, the disk has a size of 20 mas on the sky, defining the maximum distance at which it can be resolved with at least four beams.

\begin{figure}
    \centering
    \includegraphics[width=1.0\linewidth]{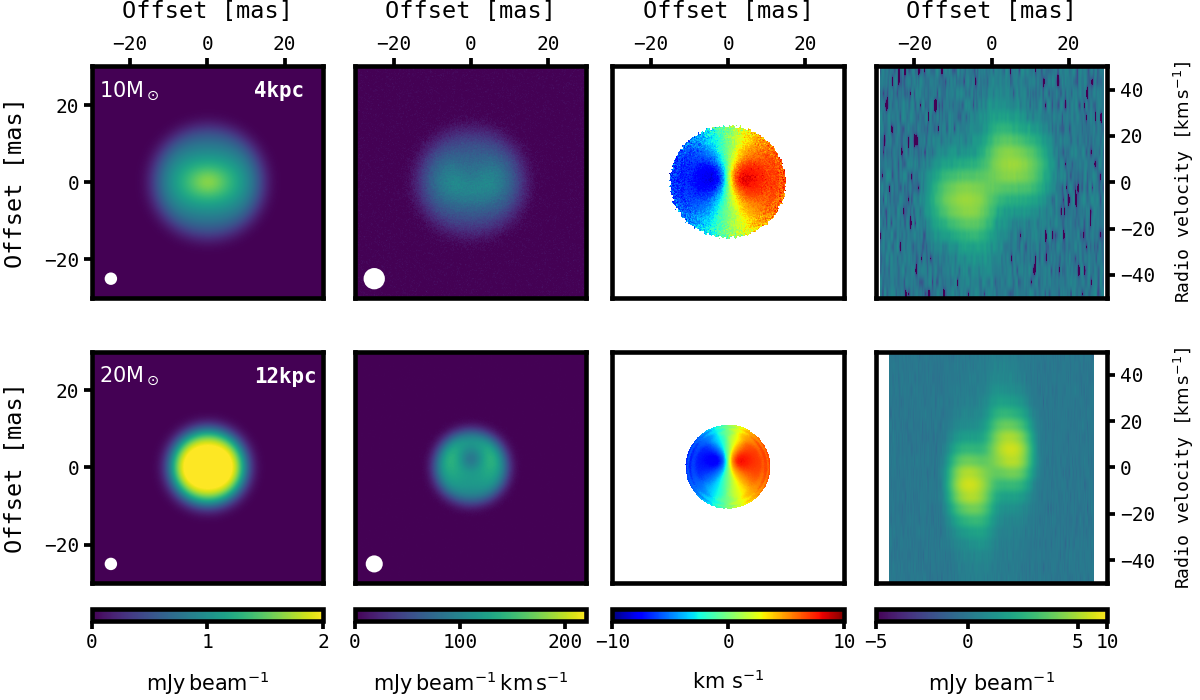}
    \caption{Simple synthetic observations of additional ionized disk models. The first row shows the observations for the 10 $\mathrm{M_\odot}$ model, while the second row shows the observation for the 20 $\mathrm{M_\odot}$ model. 
    The first column, from left to right shows the continuum emission. The second and third column shows the zeroth, and first moment maps. The fourth column shows the PV diagrams.}
    \label{fig:disk_param_models}
\end{figure}

Stars with masses below 10 $\mathrm{M_\odot}$ emit very few ionizing photons, so the faintest disks of interest for this study (those that are internally ionized) surround a 10 $\mathrm{M_\odot}$ star. The gravitational radius for such a star is 60 au, which at a distance of 4 kpc corresponds to 30 mas. Using a beam size of 4.8 mas in the synthetic observations, the disk can be resolved with approximately six beams, including three to sample the central part of the disk. Figure \ref{fig:disk_param_models} displays the resulting synthetic observations.

In both disk models, continuum emission is detected with an S/N exceeding 1000, while $\mathrm{H38\alpha}$ line emission is observed with an S/N greater than 40. This is due to the maser effect, which remains significant even in the 10 $\mathrm{M_\odot}$ model.

\bibliography{references}{}
\bibliographystyle{aasjournal}

\end{document}